\documentclass[journal]{IEEEtran}

\usepackage{ifpdf}
\usepackage{subfigure}
\usepackage{graphicx}
\usepackage{cite}
\DeclareGraphicsExtensions{.pdf,.jpg,.png}
\usepackage[cmex10]{amsmath}
\usepackage{amssymb}
\usepackage{array}
\usepackage{mdwmath}
\usepackage{mdwtab}
\usepackage{pslatex}
\usepackage{xcolor}
\usepackage{textcomp}
\usepackage{placeins}
\usepackage{eqparbox}
\usepackage{url}
\usepackage{stfloats}
\usepackage{multicol}
\usepackage{subfigure}
\usepackage{float}
\usepackage{algorithmicx}
\usepackage{algpseudocode}
\usepackage{amsthm}
\usepackage{comment}
\usepackage{balance}
\usepackage{gensymb}
\usepackage{tabularx}
\usepackage{multirow}
\usepackage{hhline}
\usepackage{cite}

\begin{document}

\title{Standardization of Extended Reality (XR) over 5G~and~5G-Advanced 3GPP New Radio}

\author{Margarita Gapeyenko, Vitaly Petrov, Stefano Paris, Andrea Marcano, and Klaus I.~Pedersen\vspace{-1mm}

\thanks{Margarita Gapeyenko, Stefano Paris, Andrea Marcano, and Klaus I.~Pedersen are with Nokia Standards, Finland, France, and Denmark. Vitaly Petrov was with Nokia Standards, Finland, and is now with Northeastern University, USA. Klaus I.~Pedersen is also with Aalborg University, Denmark.}
}


\maketitle

\begin{abstract}
Extended Reality (XR) is one of the major innovations to be introduced in 5G/5G-Advanced communication systems. A combination of augmented reality, virtual reality, and mixed reality, supplemented by cloud gaming, revisits the way how humans interact with computers, networks, and each other. However, efficient support of XR services imposes new challenges for existing and future wireless networks. This article presents a tutorial on integrating support for the XR into the 3GPP New Radio (NR), summarizing a range of activities handled within various 3GPP Service and Systems Aspects (SA) and Radio Access Networks (RAN) groups. The article also delivers a system level simulation study evaluating the performance of different XR services over NR Release 17. The paper concludes with a vision of further enhancements to better support XR in future NR releases and outlines open problems in this area.
\end{abstract}

\section{Introduction}
\label{sec:intro}

\textcolor{black}{Extended Reality (XR) has been one of the ambitious research topics under development for several decades already, covering a range of various technologies such as augmented reality (AR), virtual reality (VR), XR gaming, and others. According to~\cite{xr_devices}, there are more than 85 manufacturers and 220+ XR devices already available.} The adoption of compact wearable devices enables numerous novel use cases in consumer, industrial, and medical areas. XR also facilitates the modern global trends toward distance and hybrid workplaces, further industrial digitization, telemedicine, and collaborative online gaming, as well as lays a foundation for \textcolor{black}{Metaverse revolution~\cite{xr_vision_paper0}}.

Following high market potential, XR will inevitably play an important role in 5G-Advanced and later 6G systems~\cite{xr_vision_paper1}. Meanwhile, the successful adoption of XR requires support from the underlying wireless systems. The latter demands substantial effort from the standardization bodies, particularly, the 3rd Generation Partnership Project (3GPP). An extra challenge in supporting XR services over 3GPP New Radio (NR) cellular networks is that XR does not perfectly fit into the existing classification of fifth-generation (5G) applications and services, typically divided into enhanced mobile broadband (eMBB), ultra-reliable low latency communications (URLLC), and massive machine-type communications (mMTC)~\cite{tr_38_838}.

To better tailor further research on XR to standardized solutions, this article complements prior tutorials on XR from mainly the research angle (\cite{xr_vision_paper1,xr_vision_paper2}, among others) by summarizing and explaining the main findings from heterogeneous 3GPP activities on XR. \emph{We particularly cover how the XR use cases are classified, modeled, evaluated, and supported in state-of-the-art and next-generation cellular networks.}

{
\color{black}The key contributions of this paper are:
\begin{itemize}
\item The article provides a tutorial-in-nature summary of the major 3GPP activities in the area, the 3GPP-adopted classification of XR services, stochastic XR traffic models, and their key performance indicators (KPIs).
\item The article follows with an illustrative numerical study on the main XR-specific KPIs in realistic deployments by following the 3GPP-ratified XR evaluation methodology.
\item Finally, the prospective enhancements and research directions are outlined for better support of XR in Release~18 and beyond 5G-Advanced systems.
\end{itemize}
}


\section{Key XR Features and 3GPP Activities on XR}
\label{sec:3gpp}
By design, XR has a unique set of features. \emph{First} is a new device form factor replacing or complementing handheld smartphones with wearable head-mounted displays or smart glasses. New \textcolor{black}{devices (i.e., glasses or head mounted displays)} impose to strict requirements on user equipment (UE) power consumption, performance, and heat dissipation. Consequently, XR devices cannot do all the computations but require assistance from other nodes. \textcolor{black}{With the help of split-rendering, the multimedia generation and augmentation workload can be divided between a powerful XR server and an XR device. Most frequently, computation-intense tasks like rendering are executed on the XR server while pose correction mechanisms are executed on the XR device~\cite{VR_VDS_Stefano}.} \emph{Second}, \textcolor{black}{due to the interaction of a user with an XR application}, XR services feature specific traffic, with one or more video streams in downlink (DL) tightly synchronized with frequent motion/control updates \textcolor{black}{from a user} in the uplink (UL). \textcolor{black}{The frequency mainly depends on the user mobility and head angular speed (i.e., head rotation).} \emph{Finally}, XR is characterized by both high data rate and strict packet delay budget (PDB), placing it in between 5G eMBB and URLLC~\cite{xr_vision_paper1}.

The standardization of XR support via NR started in 3GPP back in 2016, with Service and System Aspect (SA) working group (WG) on ``Services'' specifying 5G service requirements for high-rate and low-latency XR applications~\cite{ts_22_261}. The work continued in 2018 by SA4 (``Multimedia Codecs, Systems and Services'') documenting relevant traffic characteristics in TR 26.925 and providing a survey of XR applications in~\cite{tr_26_928}. In parallel, SA (``System Architecture and Services'') standardized new 5G quality of service (QoS) identifiers (5QI) to support interactive services including XR in TS 23.501.

{
\color{black}
Recently, the SA effort was followed by 3GPP Radio Access Network (RAN) working groups in Releases\,17 and\,18. Rel.\,17 study presented a unified set of traffic models and developed the evaluation framework for XR~\cite{tr_38_838}, while studies on candidate XR solutions took place in the early stage of Rel.\,18~\cite{tr_38_835}.
} The normative Rel.\,18 work is currently ongoing in both SA and RAN, also known as the first 5G-Advanced release. The main focus of this normative work is to adapt power saving mechanisms to XR-specific traffic, further increase the number of supported XR users, and introduce XR application awareness to improve the overall performance of XR services.

\section{Selected XR Applications and Services}
\label{sec:apps}
\begin{figure}[!t]
 \centering
 \includegraphics[width=0.95\columnwidth]{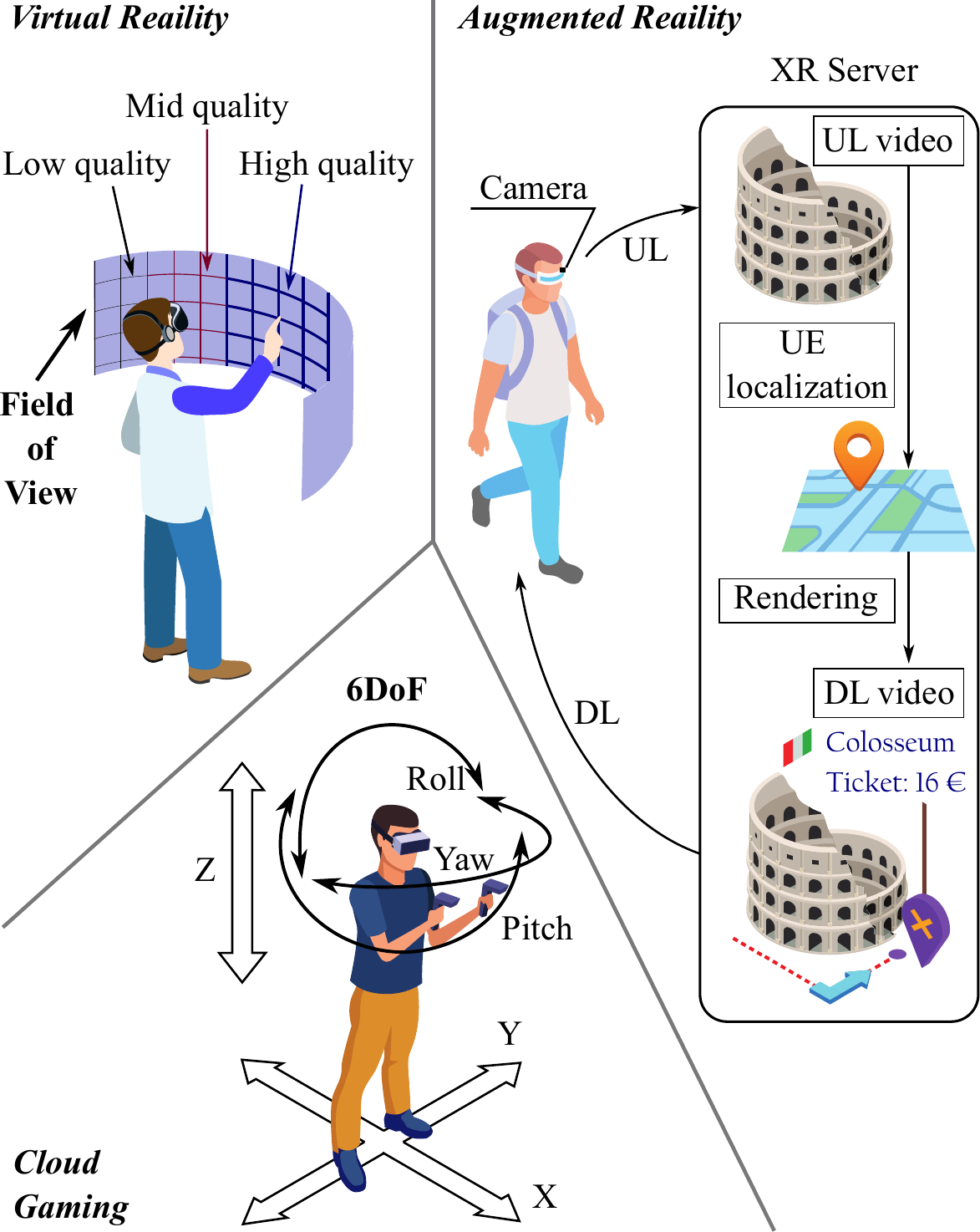}
\vspace{-2mm}
 \caption{Selected 3GPP XR use cases.}
 \label{fig:xr_usecases}
\vspace{-3mm}
\end{figure}

Decades of engineering led to the development of various XR applications and services, each characterized by its own user setup, traffic flows, and QoS metrics~\cite{xr_vision_paper1,xr_vision_paper2,xr_vision_paper_Akyildiz}. Following ~\cite{tr_26_928}, over $20$ XR use cases are identified. For such an extensive set of setups, the performance evaluation of XR wireless solutions is challenging. Therefore, it was proposed in~\cite{tr_38_838} to group the XR use cases into three meta-categories related to: virtual reality, augmented reality, and cloud gaming. We briefly introduce them below and in Fig.~\ref{fig:xr_usecases} describing the essential features to be considered in modeling.

\subsection{VR with viewport-dependent streaming}
VR generates a virtual world where a user is fully immersed, thus creating a sense of physical presence that transcends the real world. Modern VR services are usually enabled by optimized viewport-dependent streaming (VDS). VDS is an adaptive streaming scheme that adjusts the bitrate of the 3D video using both network status and user pose information~\cite{VR_VDS_han}. Specifically, the omnidirectional 3D scene with respect to the observer's position and orientation is spatially divided into independent subpictures or tiles. The streaming server offers multiple representations of the same tile by storing tiles at different qualities (varying, i.e., video resolution, compression, and frame rate). \textcolor{black}{Transmission of new XR content can be triggered by user movements, where the current user viewport is streamed with high quality while frames outside the viewport are transmitted with lower quality. The use of VDS helps to support high-rate video transmission more resource efficiently.}

\subsection{AR with simultaneous localization and mapping}
AR merges virtual objects with a live 3D view of the real world, thus creating a realistic personalized environment the user interacts with. Therefore, estimating the user location and field of view (FOV) is important. However, modern AR solutions do not rely exclusively on expensive motion detection sensors but rather complement them with cameras mounted on AR glasses. Hence, AR is often featured by a video stream in UL. The video is continuously transmitted to the XR server that performs pose tracking to estimate the position and orientation of the user via simultaneous localization and mapping (SLAM)~\cite{AR_SLAM_Stefano}.
\textcolor{black}{The AR server can be located anywhere in the network as long as the motion-to-photon latency (time between head movement is detected and a new video is displayed) does not exceed a certain delay budget. To minimize the delay budget reserved for data transmission, the XR server usually is an Edge Application Server (EAS) deployed in the Edge Data Network (EDN).}
The estimated FOV is used to generate an augmented 3D scene where virtual objects are overlaid onto certain positions. Finally, the rendered 3D objects or video are encoded and streamed back to the UE. A medium-quality UL video that captures the major objects is already sufficient for SLAM. Therefore, aiming to reduce the UL bitrate, AR UL video stream is often downscaled compared to the DL one.

{
\color{black}
\subsection{CG with 6 degrees of freedom}
Traditional gaming platforms primarily rely on a substantial part of the computing performed locally at the UE. The classic approach features low latency, especially for single-player games, but requires high-performance computing nodes deployed at the user side. These nodes (computers, game consoles, etc.) are often expensive and power-hungry, thus heating notably and also with limited mobility. 

Cloud gaming suggests replacing high-performance computing at the user terminal with a low-performance client at the user side connected to a powerful computing node in the data center or at the edge. While bringing clear benefits, CG also raises new challenges, particularly related to the latency of the connection between the UE (displaying the video and audio stream to the user and taking the user’s input) and the cloud server performing most of the computing tasks. As we further discuss, the latency boundaries are particularly stringent for interactive games involving wearable XR devices and immersive user experience. 

In a typical XR CG scenario, the server generates a sequence of 2D/3D scenes as a video stream in response to a control command sent by the UE. For XR CG, control signals include handheld controller inputs and 3 or 6 Degrees of Freedom (3DoF/6DoF) motion samples (see Fig.~\ref{fig:xr_usecases})~\cite{CG_6dof_paper}. Here, 3DoF refers to the rotation data (``roll'', ``pitch'', and ``yaw''), while 6DoF also adds the information on the UE displacement in X, Y, and Z dimensions. The resulting video stream in CG is dependent on the user's actions, so, as for VR, frequent motion/control updates are needed in UL. However, the availability of a prerecorded virtual environment at the CG server allows for slightly relaxed delay budget requirements compared to VR~\cite{tr_38_835}.
}

\section{XR-specific 3GPP Traffic Models}
\label{sec:traffic}

\begin{table*}[!t]
\vspace{-5mm}
\caption{\textcolor{black}{3GPP traffic model parameters and requirements for selected XR use cases.}}
\label{tab:xr_traffic_table}
\begin{tabular}{m{0.8cm}|m{0.4cm}|m{1.2cm}|m{1.7cm}|m{2.5cm}|m{2.55cm}|m{4.25cm}|m{0.8cm}}
\hline
Traffic stream & DL/ UL & Use cases & Packet rate & Average data rate & Packet size & Jitter & PDB\\
\hhline{========}
\multirow{4}{*}{Video} & \multirow{3}{*}{DL} & \textbf{AR} & \multirow{3}{*}{\shortstack[l]{\textbf{60\,fps},\\$[$30,90,120$]$\,fps}} & \multirow{2}{*}{\shortstack[l]{\textbf{30\,Mbit/s, 45\,Mbit/s},\\$[$$60$\,Mbit/s$]$}} & \multirow{4}{*}{\shortstack[l]{Truncated Gaussian,\\Mean = Av.\,data\,rate /\\ (fps * 8) bytes,\\$[$STD, Min, Max$]$ = \\$[$10.5\%, 50\%, 150\%$]$\\ of mean}} & \multirow{3}{*}{\shortstack[l]{Truncated Gaussian, Mean = 0\,ms\\ Standard deviation (STD) = 2\,ms\\$[$Min, Max$]$ = $[$\textbf{-4, 4}$]$\,\textbf{ms} or $[$-5, 5$]$\,ms}} & \multirow{2}{*}{$10$\,ms}\\
\cline{3-3}
& & \textbf{VR} & & & & &\\
\cline{3-3,5-5,8-8}
& & \textbf{CG} & & \textbf{8\,Mbit/s, 30\,Mbit/s}, [$45$\,Mbit/s] & & & $15$\,ms\\
\hhline{~====~==}
& \multirow{4}{*}{UL} & \multirow{2}{*}{\textbf{AR}} & $60$\,fps & \textbf{10\,Mbit/s}, \quad{}[$20$\,Mbit/s] & & Optional, same as for DL & $30$\,ms\\
\hhline{=~~=====}
\multirow{3}{*}{\shortstack[l]{Motion/\\control}} & & & \multirow{3}{*}{$250$\,fps} & \multirow{3}{*}{$0.2$\,Mbit/s} & \multirow{3}{*}{$100$\,bytes} & \multirow{4}{*}{No} & \multirow{3}{*}{$10$\,ms}\\
\cline{3-3}
& & \textbf{VR} & & & & &\\
\cline{3-3}
& & \textbf{CG} & & & & &\\
\hhline{======~=}
Audio +Data & DL+ UL & \textbf{AR/VR/CG} in DL,\newline \textbf{AR} in UL & $100$\,fps & $0.756$\,Mbit/s, $1.12$\,Mbit/s & Av.\,data\,rate / (fps*8) bytes & & $30$\,ms\\
\hhline{========}
\multicolumn{7}{l}{\emph{If multiple values given, \textbf{bold} are default values.}}
\end{tabular}
\end{table*}

The traffic model is one of the principal elements needed when simulating applications. In this section, we describe the XR traffic models for selected services. The models are based on a deep analysis of the data traces from the 3GPP trace generator and present a reasonable balance between accuracy and complexity~\cite{tr_38_838}. Apart from discussed distinct features, VR, AR, and CG have some commonalities in the employed data streams. Therefore, the description is grouped based on the data semantic, as illustrated in Table~\ref{tab:xr_traffic_table}.

\subsection{Traffic model for XR video stream}
Video stream is the flow with the highest data rate for all the considered XR use cases in DL, as well as for the AR UL. Following the traces from SA4, video is divided by a source generator into separate frames before transmission. To keep a reasonable complexity, a single data packet in the model represents multiple IP packets corresponding to the same video frame. The packet also includes the data for both left and right eyes. The packet size follows a Truncated Gaussian distribution and is determined by the average data rate in megabits per second (Mbit/s). The average inter-arrival time is an inverse of the frame rate in frames per second (fps, i.e., $60$\,fps leads to $16.6$\,ms). The actual inter-arrival time is random accounting for jitter that also follows a Truncated Gaussian distribution~\cite{tr_38_838}. Main stream parameters are summarized in Table~\ref{tab:xr_traffic_table}. \textcolor{black}{Video can be delivered as a single stream described above or via multiple streams. As per the definition in~\cite{tr_38_835}, the multi-stream approach consists of encoding several streams, each of them emphasizing a given viewport and making them available for the receiver, so that the receiver decides which stream is delivered at each time instance. One example of multi-stream model is a separate stream for left and right eyes, used to generate stereoscopic 3D video to support a close-to-reality experience.} In this case, the mean packet size is naturally reduced by 50\% and the frame rate is increased two times, compared to the baseline single-stream model. Major parameters for multi-stream video can be found in~\cite{tr_38_838}.

\subsection{Traffic model for XR motion/control stream}
The second important stream refers to motion/control updates sent by the XR device in UL, \textcolor{black}{to ensure that the latest content will be available at a UE.} 
This stream aggregates: (i) UE pose information update received from 3DoF/6DoF tracking and device sensors; and (ii) control information including user input data, auxiliary information, and/or commands from the client to the server. Following~\cite{tr_38_838}, the packet size and inter-arrival time are constant, as detailed in Table~\ref{tab:xr_traffic_table}.

\subsection{Traffic model for aggregated XR audio and data}
In addition to the video stream in UL and DL, it is possible to model audio and extra data as a separate stream. Same as for motion/control, the packet size and inter-arrival time are constant and given in Table~\ref{tab:xr_traffic_table}. According to SA4 conclusions, modeling this stream is not mandatory, as it is relatively small compared to a video. On the other side, the frame rate is higher than the DL video, which may be important for modeling certain XR power saving schemes, \textcolor{black}{as a user will need to wake up more often to receive a data.}

\bigskip

Summarizing, as illustrated in Table~\ref{tab:xr_traffic_table}, the \emph{recommended baseline setup} for the VR includes a single-stream video in DL plus a single-stream motion/control update in UL. CG employs a similar set, while the data rates and PDBs for the DL video are different. The feature of AR modeling here is the presence of UL video complementing UL motion/control updates.

\section{Major XR-specific 3GPP KPIs}
\label{sec:kpis}
 \textcolor{black}{In coherence with RAN Release 17 and 18 studies on XR~\cite{tr_38_838,tr_38_835}, we consider two major XR KPIs}: capacity and power consumption. First, a joint user-centric metric is defined for capacity and latency constraints. XR use cases are delay-sensitive: receiving a packet late has almost the same effect as losing the packet completely. So, the metric adds all the late packets to the packet error rate (PER):

\textbf{Metric 1: A satisfied XR UE}. \emph{An XR UE is declared satisfied if more than $X\%$ of application layer packets are successfully transmitted within a given PDB.} Multiple values of $X$ can be considered, while the baseline is $99\%$~\cite{tr_38_838}. This user-centric satisfaction is further extended to the network level as:

\textbf{Metric 2: XR capacity}. \emph{XR capacity is defined as the maximum number of XR UEs per cell with at least $Y\%$ of these UEs being satisfied.} The value $Y<100\%$ is used to filter out outliers in the simulations, namely unfortunate deployments with blind spots. In 3GPP RAN1 the baseline $Y$ is $90\%$~\cite{tr_38_838}.

Besides XR capacity, battery life is another important criterion determining the market potential of cellular-connected XR devices. As the absolute UE power consumption value varies a lot among device vendors, only metrics for the relative power consumption are adopted:
 
\textbf{Metric 3: UE power saving gain versus ``Always ON''}. \emph{UE power saving gain presents a ratio between the average UE power consumption when employing a certain power saving technique and the average UE power consumption when the UE continuously monitors control channels and is always available for base station (BS) scheduling.}

Most of the existing UE power saving techniques, including discontinuous reception (DRX) and physical control channel skipping (PDCCH skipping), suggest the XR UE stay in a sleep mode as long and frequent as possible. However, due to strict PDBs of XR traffic, there is not always enough time to wake up the UE, schedule the transmission, and successfully deliver the packet if it arrives during a UE sleep period. Therefore, the potential corresponding loss in XR capacity should be accounted for along with the power saving gain.

\section{XR Evaluation in NR Release 17: A Case Study}
\label{sec:results}
One of the principal research questions related to running XR over cellular networks is: \emph{``How well XR services can be supported today in a modern 5G system?''} The answer determines the state-of-the-art and can serve as an important baseline when assessing existing and future proposals for better XR support in 5G-Advanced and 6G networks.

Here, we present key system-level performance results of supporting XR services over cellular networks in realistic deployments featured by a single frequency layer 5G NR network with multiple base stations and devices. Our study follows the 3GPP evaluation methodology for ``XR over NR'' detailed in~\cite{tr_38_838,tr_38_835}.


\subsection{Simulation setup and framework}
Our numerical results are obtained via a 3GPP-calibrated Nokia proprietary system-level simulator. Key assumptions are summarized in Table~\ref{tab:params}, in line with the XR features discussed above and the 3GPP methodology~\cite{tr_38_838}. All the major functionalities of the radio access protocol stack are simulated. The adopted XR traffic models assume the XR service deployed at the edge server. The simulator operates at symbol-level (time) and sub-carrier (frequency) resolution, where the received SINR is calculated for every transmission, accounting for the effects of MIMO transmissions and receivers, spatial radio propagation effects, inter- and intra-cell interference, etc. MAC layer effects such as Hybrid ARQ and Scheduling are explicitly modeled as well. Similarly, XR frame segmentation and assembly at Packet Data Convergence Protocol (PDCP) are also explicitly modeled.

Two 3GPP deployments are modeled: (i) Dense Urban (DU) with $21$ macro cells and \textcolor{black}{(ii)} Indoor Hotspot (InH) with $12$ indoor cells~\cite{tr_38_838}. \textcolor{black}{The InH encompasses a building size of 50x120 meters, while the DU area of 21 cells measures 530x460 meters~\cite{tr_38_838}.}
In DU, $20\%$ of UEs are outdoor, the rest are distributed indoors across floors 1--8. We assume even-load random placement of XR UEs, where the UEs perform cell selection based on reference signal received power (RSRP). We study both NR frequency ranges (FRs): sub-6GHz FR1 and millimeter wave FR2 at 30GHz. We particularly focus on baseline XR setups with the XR DL video stream complemented by pose updates in the UL following the characteristics discussed earlier. 


\begin{table}[!b]
\caption{Case study simulation parameters.}
\label{tab:params}
\vspace{-3mm}
\begin{tabular}{m{3.6cm}|m{1.8cm}|m{1.8cm}}
\hline
\multicolumn{3}{c}{\textbf{\emph{General}}}\\
\hline
Channel state information (CSI)& \multicolumn{2}{c}{Periodic every $2$\,ms}\\
\hline
Channel estimation& \multicolumn{2}{c}{Realistic with ideal CSI}\\
\hline
Modulation and coding (MCS)& \multicolumn{2}{c}{Up to 256-QAM}\\
\hline
Frame structure& \multicolumn{2}{c}{DDDSU}\\
\hline
Scheduler (single-user MIMO)& \multicolumn{2}{c}{Proportional fairness}\\
\hline
Target block error rate& \multicolumn{2}{c}{$10\%$ for the first transmission}\\
\hline
UE speed& \multicolumn{2}{c}{$3$\,km/h}\\
\hline
\multicolumn{3}{c}{\textbf{\emph{Deployments}}}\\
\hline
 & \quad{}\quad{}\,\,\textbf{DU} & \quad{}\quad{}\,\,\textbf{InH}\\
\hline
Channel model & UMa & InH\\
\hline
Inter-site distance & $200$\,m & $20$\,m\\
\hline
Base station (BS) height & $25$\,m & $3$\,m\\
\hline
Antenna downtilt & $12\degree$ & $90\degree$\\
\hline
\multicolumn{3}{c}{\textbf{\emph{Radio}}}\\
\hline
 & \quad{}\quad{}\,\,\textbf{FR1} & \quad{}\quad{}\,\,\textbf{FR2}\\
\hline
Carrier frequency & $4$\,GHz & $30$\,GHz\\
\hline
Subcarrier spacing & $30$\,kHz & $120$\,kHz\\
\hline
System bandwidth & \multicolumn{2}{c}{$100$\,MHz}\\
\hline
BS noise figure & $5$\,dB & $7$\,dB\\
\hline
UE noise figure & $9$\,dB & $13$\,dB\\
\hline
BS antenna & $32$\,TxRUs, $5$\,dBi gain & \textcolor{black}{$2$\,TxRUs with grid of beams, $5$\,dBi gain}\\
\hline
UE antenna & 2T/4R, \newline \textcolor{black}{$0$\,dBi gain} & \textcolor{black}{3 panels (left, right, top), $5$\,dBi gain}\\
\hline
BS Tx power & \textbf{DU}: $51$\,dBm\newline \textbf{InH}: $31$\,dBm & \textbf{DU}: $51$\,dBm\newline \textbf{InH}: $24$\,dBm\\
\hline
UE Tx power & \multicolumn{2}{c}{$23$\,dBm}\\
\hline
\end{tabular}
\end{table}

\subsection{Capacity evaluation}

\begin{figure}[!t]
 \centering
 \includegraphics[width=0.95\columnwidth]{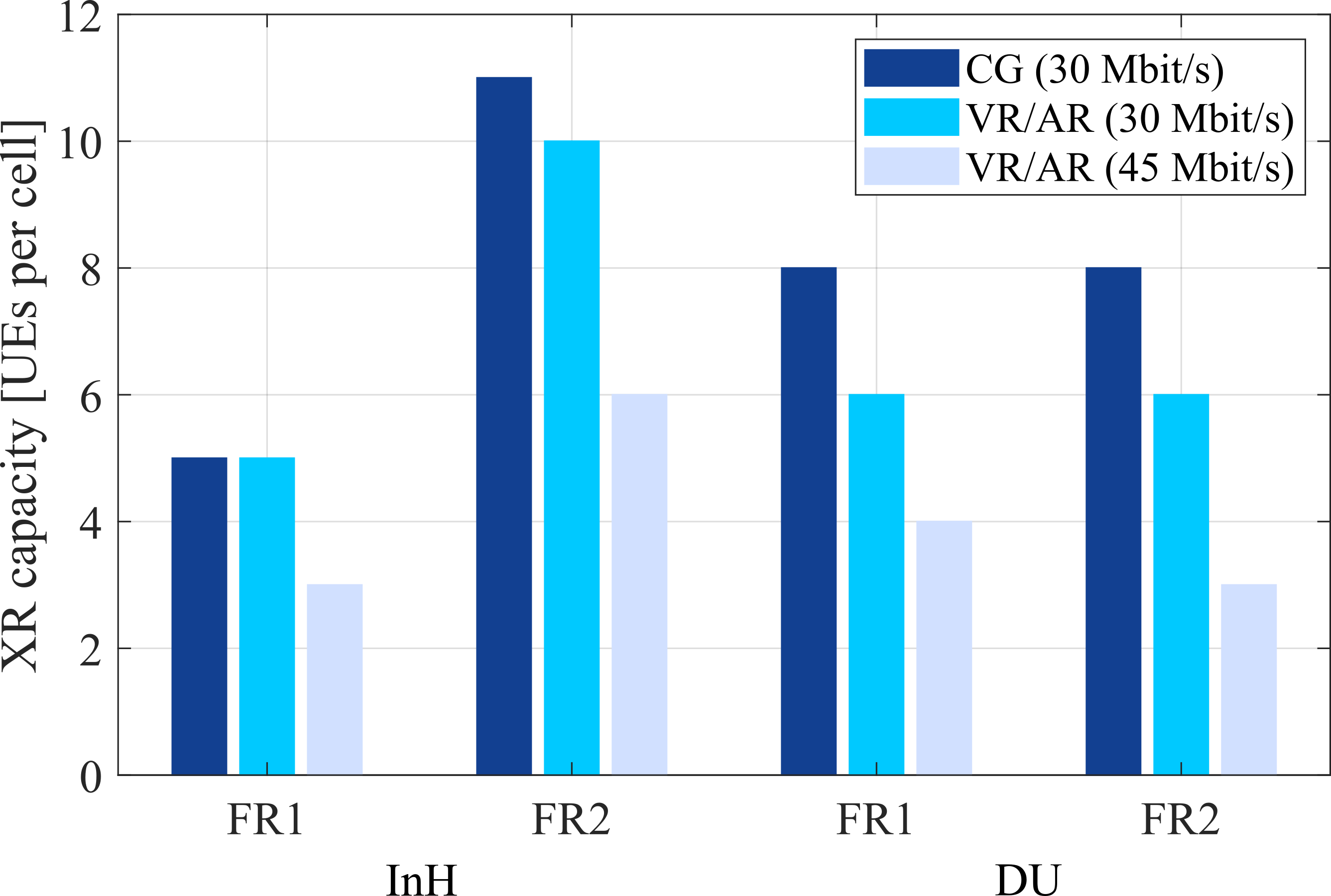}
\vspace{-2mm}
 \caption{XR capacity evaluation results in FR1 and FR2.}
 \label{fig:xr_capacity}
\end{figure}

Fig.~\ref{fig:xr_capacity} illustrates the XR network capacity (minimum of DL and UL) achievable for typical XR setups in InH and DU deployments with either FR1 or FR2 radio. We first note that the considered XR use cases are DL-limited, not UL-limited, as our study shows that more XR devices per cell can be supported in UL than in DL in every modeled configuration. Hence, DL video is a major factor limiting XR network capacity in Fig.~\ref{fig:xr_capacity}. Particularly, only six XR devices are supported per cell for ``AR/VR 30\,Mbit/s'' in DU FR1. The limitation here is the stringent PDB requirement of $10$\,ms imposed for AR/VR services (see Table~\ref{tab:xr_traffic_table}). Meanwhile, CG with $15$\,ms PDB features up to $1.3$ times higher XR capacity for the same $30$\,Mbit/s per UE. The XR capacity for CG ultimately reaches $11$ UEs per FR2 cell in InH. We finally observe that the change of $30$\,Mbit/s with $45$\,Mbit/s decreases the XR capacity by up to $2$\,times, so XR capacity scales non-linearly with the rate, as latency plays a critical role.

\subsection{Power consumption evaluation}
\begin{figure}[!t]
 \centering
 \includegraphics[width=0.95\columnwidth]{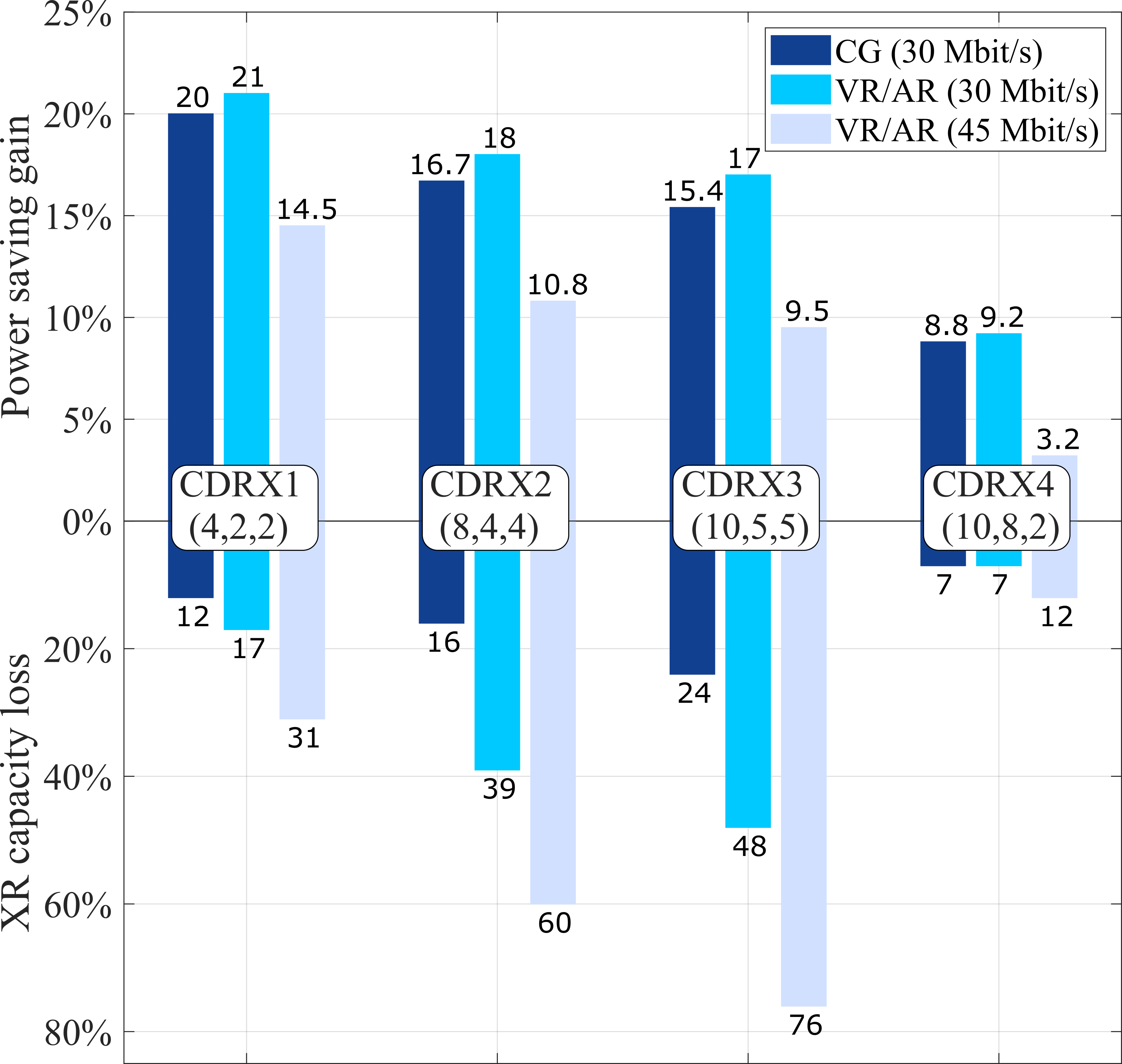}
 \caption{XR power saving gain evaluation results in FR1 DU.}
 \label{fig:xr_power}
\vspace{-5mm}
\end{figure}

Fig.~\ref{fig:xr_power} shows the DL capacity and XR UE power consumption in FR1 DU with different power saving techniques. We model four connected mode DRX (CDRX) configurations, where the values in brackets \textcolor{black}{in Fig.~\ref{fig:xr_power} (e.g., CDRX1 (4, 2, 2)) stand for: ``DRX long cycle (ms)'', ``On duration timer value (ms)'', and ``Inactivity timer value (ms)''.} \textcolor{black}{Following Release 17 XR power saving methodology}, the gains are calculated versus ``UE Always ON'' scheme, where UE is always available for scheduling~\cite{tr_38_838}. From Fig.~\ref{fig:xr_power} we first observe that the increase of the ``On duration'' decreases the power saving gain. After a short ``On duration'', there is a high probability of going to sleep, as the chances of immediately receiving the next video frame are low.

\textcolor{black}{Comparing CDRX3 (10\,ms, 5\,ms, 5\,ms) with CRDX4 (10\,ms, 8\,ms, 2\,ms), we note that the capacity loss decreases for the same cycle duration and increased ``On duration'' from 5\,ms to 8\,ms.} The latter is caused by the fact that higher ``On duration'' increases the probability of receiving the video frame before its PDB expires. We finally notice that ``AR/VR 45Mbit/s'' is the most challenging setup, featuring the lowest power saving gain and highest capacity loss. Here, reception of larger video frames takes a longer time, which both compromises strict PDB of $10$\,ms and demands the XR UE to stay awake longer once the transmission is scheduled.

\section{Outlook on 5G-Advanced innovations for XR}
\label{sec:challenges}

\begin{figure*}[!t]
 \centering
\vspace{-2mm}
 \includegraphics[width=0.95\textwidth]{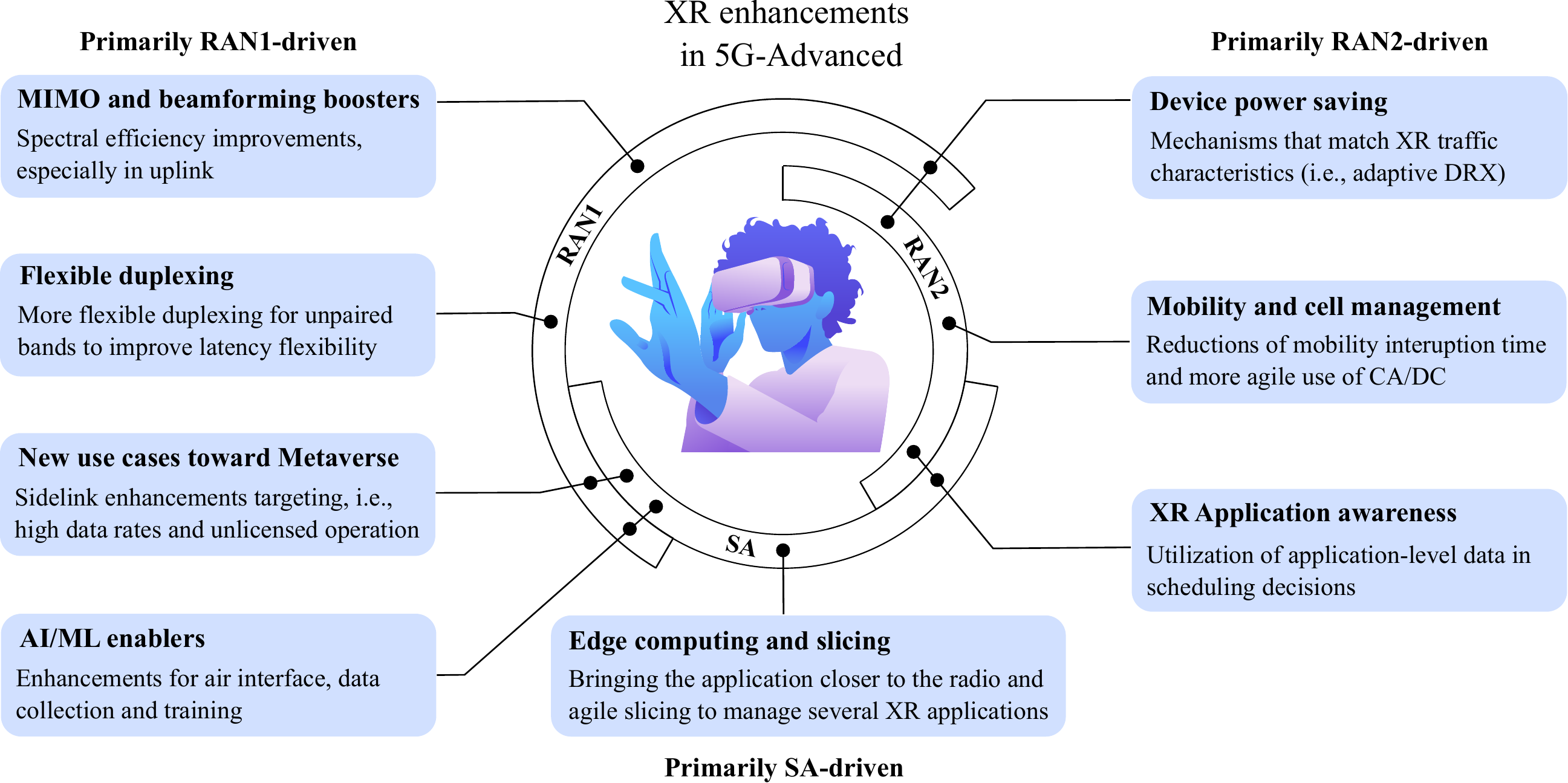}
\vspace{-2mm}
 \caption{\textcolor{black}{Improving support for the XR in 5G-Advanced.}}
 \label{fig:xr_5g_advanced}
\end{figure*}

Although legacy 5G is proven capable of providing basic XR support, massive adoption of XR devices and services in cellular networks imposes new research and engineering challenges. To further boost the XR performance, 5G-Advanced (starting with NR Release 18) will introduce a multitude of heterogeneous enhancements, including both XR-specific enablers and service-agnostic enhancements that will also improve the XR performance. An overview of these innovations is given in the following and illustrated in Fig.~\ref{fig:xr_5g_advanced}.

\subsubsection{XR application awareness and scheduling}
Among the main XR-specific enhancements, a higher degree of application awareness is to be introduced, thus enabling more efficient Radio Resource Management (RRM) and scheduling. This involves innovations for both RAN and SA toward end-to-end optimized XR solutions for advanced multi-modality services supporting efficient interaction between 5G system and application. For example, packet data unit (PDU) Set identifiers can be used to differentiate network handling of packets originated when an application-layer data unit is fragmented. The PDU Set identifier can be carried though the 5G network as an extension to the data plane header (GTP-U) or using dedicated control plane messages. In addition, new metrics related to application-layer QoS requirements and new assistance information signaled by the core network (CN) to the RAN have been specified for interactive XR and gaming services. \textcolor{black}{Examples include Low Latency, Low Loss, Scalable Throughput (L4S) schemes in the 3GPP RAN and CN, which is being further enhanced for use cases with XR services.} Furthermore, scheduling enhancements improving buffer status report and configured grant for UL are expected to better fit the time-variant XR traffic and stringent latency requirements. \textcolor{black}{The configured grant enhancements include: (i) multiple transmission occasions in a period to support large XR payloads and (ii) dynamic indication of unused configured grant resources to deal with XR payload variations~\cite{tr_38_835}.} Also packet discarding mechanism (dropping late XR packets that will not meet the latency requirements anyway) can further facilitate higher XR capacity.  

\subsubsection{XR device power saving}
Dedicated power saving optimizations for XR devices are also envisioned. These are needed to efficiently minimize battery draining and dissipated heat, which is a major concern for users’ comfort if using XR glasses. To address these challenges, adaptive solutions are to be further developed that autonomously learn the best configuration for existing power saving techniques, including DRX, sparse PDCCH monitoring, and Wake Up Signal (WUS)~\cite{klaus_3}. For example, parameters of the DRX duty cycle and the frequency for monitoring the PDCCH may be dynamically adjusted according to the XR traffic characteristics, XR requirements, network load, and radio channel status. \textcolor{black}{Additionally, novel schemes to compensate time drift due to the mismatch between DRX duty cycle and the period of XR traffic have been introduced in the latest release~\cite{tr_38_835}.} 

\subsubsection{MIMO and beamforming}
While 5G already supports advanced multiple-input multiple-output (MIMO) and beamforming, 5G-Advanced is particularly focused on UL enhancements, such as higher ranks and more transmission chains that could bring an estimated additional 20\% gain in UL throughput gain. This will offer benefits, especially for UL-heavy AR applications. In addition, optimizations for dynamic beam management will also be considered.

\subsubsection{Mobility and cell management}
Mobility-centric innovations are also on the 5G-Advanced agenda. From an XR perspective, the desire to further lower service interruption time during handover is particularly important. Today, the handover interruption time is around 20-50 ms for basic and conditional handovers and down to 2-10 ms for dual active protocol stack (DAPS) handovers. Such interruption times can be harmful to the user experience if the XR UE is subject to frequent handovers. The intention is, therefore, to lower these values for 5G-Advanced, aiming at truly seamless handovers for the UEs, including devices with only one transmitter/receiver chain and protocol stack (without DAPS support). Further gains can be obtained from advanced carrier aggregation and dual connectivity (CA/DC) mechanisms.

\subsubsection{Edge computing and slicing}
The trend to introduce edge computing enhancements will continue in the 5G-Advanced era, particularly toward bringing the Application Server (AS) on the network side closer to the application client on the UE side. While legacy 5G already includes several edge computing features, it is foreseen that 5G-Advanced may also incorporate, among others, different policies for different categories of UEs, as well as the ability to relocate a given edge application server for a collection of UEs. Here, XR is one of the major use cases for edge computing due to its stringent latency requirements. Network slicing is another useful mechanism to manage highly diverse XR services over a single 5G network.

\subsubsection{Sidelink enhancements and new use cases}
A plethora of sidelink enhancements are also considered for 5G-Advanced. Among others, those include carrier aggregation enhancements to support higher data rates and improved reliability, as well as sidelink operation for unlicensed bands. Such sidelink enhancements could enable new XR applications, i.e., between smartphones or smart glasses, ultimately paving the way toward the Metaverse revolution~\cite{xr_vision_paper1}.

\subsubsection{Flexible duplexing}
5G-Advanced will also study flexible duplexing (FDU) evolution, including enabling simultaneous UL and DL transmission on an unpaired carrier. FDU is envisioned to offer additional flexibility for latency-throughput demanding XR applications, as compared to the current TDD solution with either exclusive DL or exclusive UL transmission at any given time for each unpaired carrier per cell~\cite{xr_fdd}. Here, the feasibility and cost of self-interference schemes to enable simultaneous DL and UL transmission on an unpaired carrier is of primary importance.

\textcolor{black}{
\subsubsection{AI/ML enablers}
Functionalities to fully enable the use of Artificial Intelligence (AI) and Machine Learning (ML) techniques are also rapidly coming with 5G-Advanced. AI/ML techniques are considered of paramount importance for operating networks optimally for XR services. AI/ML enablers are therefore being introduced at all layers of the system. As a few examples, Release 18 is currently conducting a study of AI/ML enhancements for the lower layers of the air interface, while higher layer AI/ML techniques for the RAN and SA have been gradually developed since Release 16, lately also including more options for operations like data collection and training. For example, AI/ML can be used for learning XR traffic patterns, implementing smarter application-aware scheduling~\cite{xr_vision_paper_Akyildiz}, and deploying intelligent configuration tools for DRX to maximize capacity and energy savings.
}

\section{Conclusions and The Road Ahead}
\label{sec:conclusions}

Reflecting major 3GPP activities on XR, this article provides an executive summary on the XR via NR in 5G and 5G-Advanced. A case study is also presented confirming that 5G NR can already support XR services, albeit with a limited number of XR devices per cell for the highest data rates.

Massive adoption of XR devices and services is one of the drivers for the development of future cellular systems. Further XR-related innovations are likely to appear in upcoming NR releases. Among many others, this may include enhanced UE positioning and orientation via network localization and sensing techniques as well as the adoption of AI/ML to better adapt to time-variant XR traffic. Hence, while state-of-the-art 5G networks can already run some XR services, the standardization work on better XR support is still in its early stage, with many open research problems to be tackled for 5G-Advanced and 6G systems.

\section*{Acknowledgment}
The authors are grateful to Rastin Pries, Mads Lauridsen, Ece Ozturk, Karri Ranta-Aho, Devaki Chandramouli, and Antti Toskala from Nokia for their useful suggestions and feedback. 


\balance

\section*{Authors' Biographies}

\textbf{Margarita Gapeyenko} (margarita.gapeyenko@nokia.com) is a Senior Standardization Specialist and a 3GPP RAN1 Delegate at Nokia. She received her M.Sc. from the University of Vaasa, Finland, in 2014 and her Ph.D. from Tampere University, Finland, in 2022. Her research interests include millimeter wave networks, UAV communications, 5G-Advanced, and 6G wireless systems.

\textbf{Vitaly Petrov} (v.petrov@northeastern.edu) is an Associate Research Scientist, Northeastern University, Boston, USA. Earlier, he was a Senior Standardization Specialist and a 3GPP RAN1 Delegate at Nokia. Vitaly received his Ph.D. from Tampere University, Finland. His research interests are in millimeter wave and terahertz band networks.

\textbf{Stefano Paris} (stefano.paris@nokia.com) is a Senior Research Scientist at Nokia Standards. He received his B.Sc. (2004) and M.Sc. (2007) from University of Bergamo, and his Ph.D. (2011) from Politecnico di Milano. His research interests include resource allocation, optimization, artificial intelligence and machine learning applied to wireless and wired networks.

\textbf{Andrea Marcano} (andrea.marcano@nokia.com) received her B.Sc. (2011) from the Católica Andrés Bello University and M.Sc. (2013) from the Simón Bolívar University, Venezuela; her Ph.D. degree (2018) is from Technical University of Denmark. She is a Research and Standardization Engineer at Nokia, France. Her research interests are in 5G NR networks.

\textbf{Klaus I. Pedersen} (klaus.pedersen@nokia.com) received his M.Sc. degree (1996) and his Ph.D. (2000) from Aalborg University, Denmark. He is a Nokia Bell Labs Fellow, leading the Radio Access Research Team in Aalborg, and an external professor at Aalborg University. His research covers access protocols and radio resource management for 5G-Advanced.
\end{document}